\documentclass[11pt,a4paper]{article}
\usepackage{jinstpub}
\usepackage[utf8]{inputenc}
\usepackage{subfig}
\usepackage{lineno}

\title{The Calibration Units of KM3NeT}

\subheader{Very Large Volume Neutrino Telescope 2021 conference\\18-21 May, 2021\\Valencia / online}

\author[a]{R. Le Breton\footnote{Corresponding author}}
\author[b]{, M. Billault}
\author[a]{, C. Boutonnet}
\author[a]{, C. Champion}
\author[a]{, S. Colonges}
\author[b]{, A. Cosquer}
\author[a,f]{, A. Creusot}
\author[b]{, S. Henry}
\author[a]{, A. Ilioni}
\author[b]{, P. Keller}
\author[b]{, P. Lagier}
\author[c]{, R. Lahmann}
\author[b]{, P. Lamare}
\author[a]{, J. Lesrel}
\author[a]{, M. Lindsey Clark}
\author[b]{, J. Royon}
\author[d]{, G. Riccobene}
\author[e,g]{, D. Samtleben}
\author[a,f]{, V. Van Elewyck and}
\author[]{ on behalf the KM3NeT Collaboration}

\emailAdd{remy.lebreton@apc.in2p3.fr}

\affiliation[a]{Université de Paris, CNRS, Astroparticule et Cosmologie, F-75013 Paris, France}
\affiliation[b]{Aix Marseille Univ, CNRS/IN2P3, CPPM, Marseille, France}
\affiliation[c]{Friedrich-Alexander-Universitat Erlangen-Nurnberg, Erlangen Centre for Astroparticle Physics, Erwin-Rommel-Strasse 1, 91058}
\affiliation[d]{INFN, Laboratori Nazionali del Sud, Via S. Sofia 62, Catania, 95123 Italy}
\affiliation[e]{Nikhef, National Institute for Subatomic Physics, Amsterdam, Netherlands}
\affiliation[f]{Institut Universitaire de France, 1 rue Descartes, Paris, 75005 France}
\affiliation[g]{Leiden University, Leiden Institute of Physics, PO Box 9504, Leiden, 2300 RA Netherlands}

\abstract{KM3NeT is a deep-sea infrastructure composed of two neutrino telescopes being deployed in the Mediterranean Sea: ARCA, near Sicily in Italy, designed for neutrino astronomy, and ORCA, near Toulon in France, designed for neutrino oscillation physics. To achieve the best performance, the exact location of the optical modules, affected by sea current, must be known at any time and the timing resolution between optical modules must reach the nanosecond. Moreover, the properties of the environment in which the telescopes are deployed must be continuously monitored because they affect the timing and positioning calibration. KM3NeT is going to deploy several dedicated Calibration Units to meet these calibration goals. Because of the difference in size between ARCA and ORCA, the design of the Calibration Unit is not the same for the two sites. This proceeding describes all the devices, features and purposes of the Calibration Units with a focus on the ORCA Calibration Unit.}

\keywords{Detector alignment and calibration methods (lasers, sources, particle-beams), Instrument optimisation, Cherenkov detectors, Neutrino detectors, KM3NeT.}

\date{18-21 June 2021}

\begin{document}

\linenumbers

\maketitle

\section*{Introduction}

The KM3NeT \cite{km3net} infrastructure is composed of two telescopes, ORCA and ARCA. They are built with the same technology, based on the so-called Digital Optical Module (DOM) hosting 31 photomultipliers, but they have different geometries in order to probe different ranges of neutrino energy: GeV scale for ORCA and TeV to PeV for ARCA. Arranged in a 3D array (with 18 DOMs per Detection Unit \cite{domdu}), these DOMs can detect Cherenkov radiation that can be the signature of a neutrino interaction, from which the energy and direction of the incoming particle can be reconstructed. To properly reach the required angular resolution, all elements of the telescope need to be calibrated to ns level timing accuracy \cite{timecalib}, and in position with an accuracy of about 10 cm \cite{poscalib}. Moreover, water properties have an impact on light and sound propagation, which are two crucial parameters for calibrations. The evolution of environmental conditions must be monitored, that is mainly why Calibrations Units (CU) \cite{cubefore} will be deployed in the vicinity of the telescopes.

\section{Calibration Units}

\begin{figure}[!ht]
\centering
\includegraphics[width=.5\textwidth]{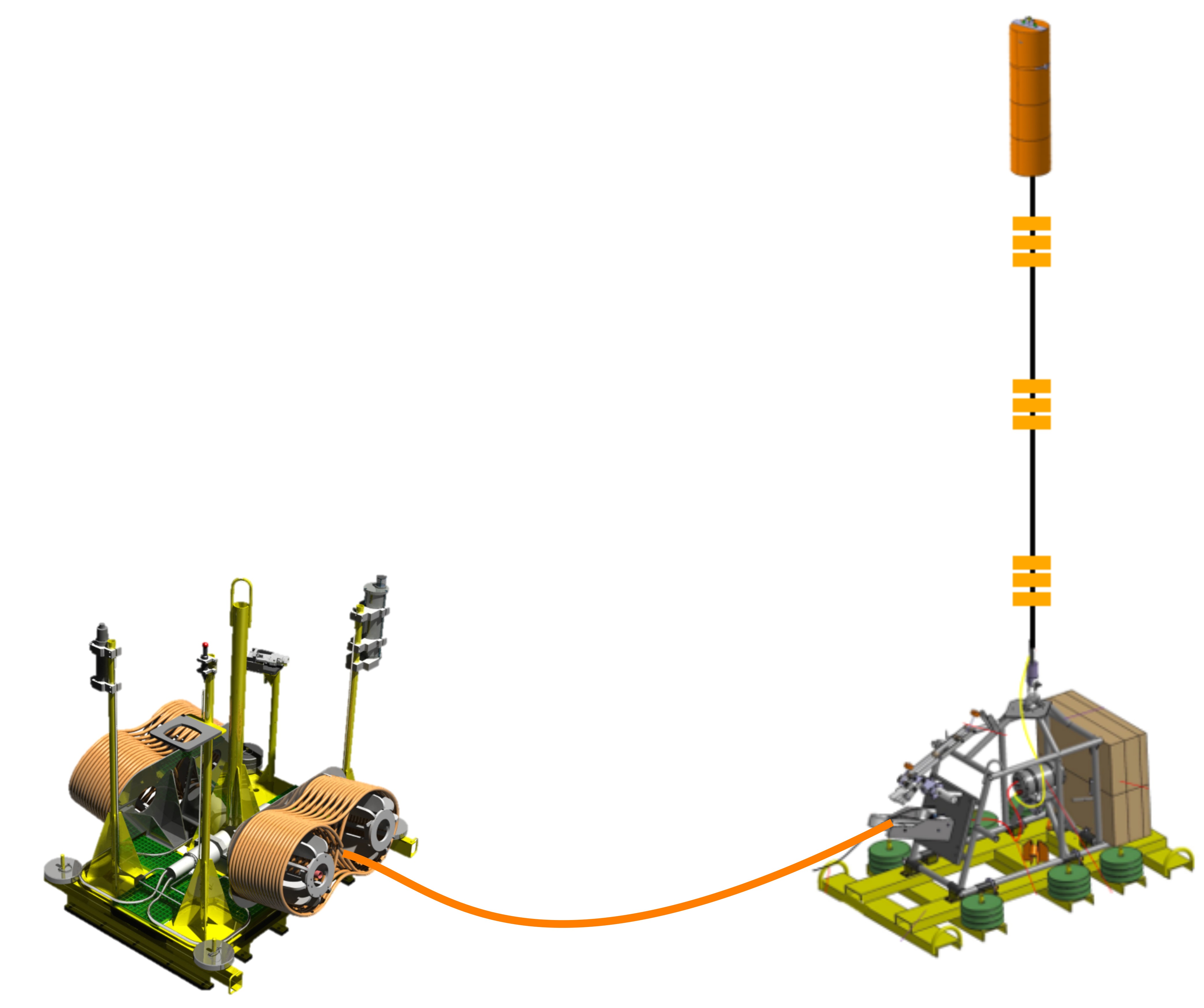}
\caption{General design of the KM3NeT Calibration Units. The Calibration Base is on the left, and is linked to Instrumentation Unit, on the right. (Not to scale)}
\label{fig:cufull}
\end{figure}

The CUs are composed of a Calibration Base and an Instrumentation Unit (figure \ref{fig:cufull}). The Instrumentation Unit is connected to the Calibration Base by a wet-mateable interlink cable, about 50 m apart for ORCA. The Calibration Base hosts the calibration devices on dedicated masts: an acoustic emitter and receiver (positioning calibration), and, in the case of ORCA, a Laser Beacon (time calibration). The Instrumentation Unit is composed of an Instrumentation Base and an inductive, semi-autonomous and recoverable Instrumentation Line.

\subsection{ORCA Calibration Base}
\label{sec:cb}

Because of deployment constraints, the Calibration Base will be deployed 40 m from the ORCA array of Detection Units. The scientific goal of the Calibration Base is to participate in the time synchronisation between Detection Units, thanks to the Laser Beacon. It will also be part of the long baseline acoustic positioning system thanks to the addition of a hydrophone and a permanent acoustic emitter. The main components of the Calibration Base are described in this section.

The steel anchor is painted with a yellow water-resistant paint. It has three masts to host the instruments for calibrations. The anchor also hosts the interlink cables, one that is going to the Instrumentation Unit, and the other to a Detection Unit. Anodes for protection against corrosion have been added on the bottom sides. A special part has been designed to host one end of the interlink cable so that the remotely operated vehicle that is used for the submarine connections can grab it more easily. The Base Module is located below the main lifting mast in the middle. The centre of gravity can be adjusted to be on the centre of the whole structure thanks to weights.

The Base Module is where the optical connections between a Detection Unit, the Calibration Base and the Instrumentation Unit are handled. A power board is used to control and dispatch power to all the instruments. All the commands are handled by a Central Logical Board, with a FPGA Mezzanine Board where the instruments and Instrumentation Unit are plugged in.

The hydrophone, (Colmar \cite{colmar}, double gain, DG330), is used to compute the position of the Calibration Base on the seabed. It can also be used for cetaceans monitoring. Data are sent to the Central Logical Board for time-stamping and transmission to shore. The connection is made by a RJ45 connector through which the power, data and clock signals are routed from/to the hydrophone. Its sampling frequency is 195.3 kHz, and it can record signals from 5 kHz to 90 kHz.

The acoustic emitter is a commercial item, supplied by Mediterraneo Senales Maritimas \cite{msm} (MAB 100, Ti Body, Gisma connector), and used for DOM positioning. The emitter has its own modulation signature carried by a signal ranging from 10 kHz to 40 kHz. The acoustic sensors on the DOMs can then detect unambiguously signals from different emitters at the same time. The core of the device is a FFR SX30 acoustic transducer, and the electronic boards are hosted in a shielded container resistant up to 400 bars (ORCA is operating at 250 bars, ARCA at 350 bars). The emitter is connected to the Calibration Base container by MCIL6F/MCIL6M connectors. Communications and triggers synchronised with the detector master clock are handled through a RS232 line. 

The Laser Beacon will perform time calibration as well as water properties measurements. It is based on a commercial Nd-YAG pulsed laser (STG-03E-140 and MLC-03A-BP1 control board, Teem Photonics \cite{teem}) with sub nanosecond pulses (3.8 $\mu$J, 0.4 ns FWHM). The frequency of the pulses can be changed from 1 Hz to 4 kHz. In order to artificially modify the energy of the pulses, an optical attenuator (LVA-100-VIS, Meadowlark Optics \cite{meadow}) is installed just after the exit of the laser. Then the light goes through a diffuser (OPAL diffuser 10DIFF-VIS from Newport \cite{newport}) in order to have a maximum amount of light leaving the Laser Beacon by the vertical sides of an optical rod in borosilicate, located on one end of the LB. Simulations were made to validate the design of this rod, and check the performances of the LB. The current needed for the laser operation cannot be drawn directly from the main power supply. This is why a nickel-metal hybrid rechargeable battery (from RS \cite{rs}, Ni-MH HTD cells connected in series) has been added to the system, to be used as a local power accumulator. Is has a typical 7000 mAh capacity, a 0.1 CmA charge and a 0.2 CmA discharge rates. It is foreseen to use the Laser Beacon from once a month up to once a week. A custom electronics card, the Laser Power Management and Interface, has been designed to allow the control of the Laser Beacon and the battery. It communicates with the FPGA Mezzanine Board in the Base Module through a RS232 connection.

\subsection{ORCA Instrumentation Unit}
\label{sec:iu}

The Instrumentation Unit will monitor the water properties along the water column of ORCA (200 m), in order to compute the speed of sound in water, a crucial parameter for positioning calibration.

The Instrumentation Base titanium container is connected to the Calibration Base by a 50 m RS422 link. It hosts: an Inductive Modem Module (Seabird \cite{seabird}) using Differential Phase-Shift Keying to allow data transmission with low error rates; a RS422 to RS232 converter; a DC/DC converter to adapt the 12V sent by the Calibration Base; one penetrator for the connectivity toward the Inductive Cable Coupler from Seabird, which transforms the data from inductive modulation in the cable into an electrical signal.

The Inductive Cable Coupler links the inductive cable of the Instrumentation Line and the Instrumentation Base. The inductive cable is made of a steel wire protected by a polypropylene layer, except on its ends to allow grounding with seawater. It is kept vertical thanks to a buoy in synthetic foam. The Inductive Modem Module couples inductively to the cable along the insulated part of the cable, without direct electrical connection, thanks to the Inductive Cable Coupler. The inductive cable allows for only one current path, so data can only be polled sequentially from one instrument at a time. The Instrumentation Line hosts three main commercial probes, replicated at three different elevations: A probe (SBE SMP CTD device from Seabird) to measure conductivity, temperature and depth to compute the sound velocity as a function of temperature, pressure and salinity thanks to the seawater equation of state; A sound velocimeter (Mini SVS from Valeport \cite{valeport}) - This instrument has no native inductive interface, so to sent data through the inductive line, it is interfaced with the latter thanks to an RS232-inductive link; A current meter (AQUADOPP from Nortek \cite{nortek}). Instruments and batteries need to be re-calibrated or replaced every couple of years, this is why the Instrumentation Unit has been designed to be recoverable.

\section*{Conclusion and Outlook}

All the parts of the ORCA CU, the firmware, software and GUI are in final configuration and under tests. The final integration and tests between the Calibration Base and the Instrumentation Unit will start during summer 2021, and the deployment of the CU is foreseen in the second half of 2021.

\section*{Acknowledgement}

The authors gratefully acknowledge the support of LabEx UnivEarthS (ANR-10-LABX-0023 and ANR-18-IDEX-0001) for their research.

\end{document}